%% file: main.tex
\pdfoutput=1
\documentclass[10pt,twocolumn,letterpaper]{article}

\usepackage{cvpr}

\usepackage{graphicx}
\usepackage{amsmath}
\usepackage{amssymb}
\usepackage{booktabs}
\usepackage{multirow}
\usepackage{array}
\usepackage{xcolor}
\usepackage{pifont}
\usepackage{tikz}
\usetikzlibrary{arrows.meta,positioning,fit,backgrounds,calc}
\definecolor{cvprblue}{rgb}{0.21,0.49,0.74}
\usepackage[pagebackref,breaklinks,colorlinks,allcolors=cvprblue]{hyperref}

\newcommand{\wall}{\ensuremath{T_{\text{win}}}}
\newcommand{\real}{\ensuremath{T_{\text{rt}}}}
\providecommand{\SI}[2]{#1\,#2}

\def\confName{CVPR}
\def\confYear{2026}

\begin{document}

\title{ReGenVC: End-to-End Real-Time Generative Video Coding\\
at Ultra-Low Bitrate}

\author{Zheyuan Zhang \qquad Johnson Wu\\
Intel Corporation\\
{\tt\small \{zheyuan.zhang, johnson.wu\}@intel.com}
}

\maketitle

\input{sec/0_abstract}
\input{sec/1_intro}
\input{sec/2_related}
\input{sec/3_method}
\input{sec/4_timing}
\input{sec/5_experiments}
\input{sec/6_conclusion}

{
\small
\bibliographystyle{ieeenat_fullname}
\bibliography{references}
}

\end{document}

%% file: sec/0_abstract.tex
\begin{abstract}
We present ReGenVC (Real-time Generative Video Coding), an end-to-end \emph{generative video codec} that compresses
talking-head video to an ultra-low bitrate and decodes it in real time. On the
encoder side a source clip is reduced to a compact bitstream---a neurally
compressed first frame, per-frame pose keypoints, and metadata---totalling only
$\sim$\SI{26}{kB} for a 77-frame sequence. On the decoder side a four-step
distilled diffusion transformer reconstructs the video conditioned on the
transmitted pose and reference frame. Against x264/x265, ReGenVC transmits the same clip at
roughly \textbf{one order of magnitude lower bitrate}
($\sim$\SI{26}{kB} vs.\ the $\sim$250--280\,kB x264/x265 need for
essentially artifact-free reconstruction); at a matched ultra-low bitrate the
traditional codecs collapse into blocking artifacts while ReGenVC stays sharp by
exploiting a strong generative prior. The central obstacle to deploying such a
codec is \emph{decoder latency}: a multi-step sampler over a transformer and a
VAE is far too slow for interactive use. We make the decoder \emph{real-time} through
four-step distillation of the teacher and three model-preserving systems
techniques: (i)~eight-GPU unified sequence parallelism (Ulysses$\times$Ring),
(ii)~a spatially-split VAE, and (iii)~a three-stage overlapped pipeline; an analytical timing model
$\wall\approx(w_d{+}d)+\max(v{+}a,w_e{+}e)$ characterizes the real-time
feasibility region. On an 8-GPU node the system sustains
\SI{24}{fps} output (\SI{972}{ms} per 25-frame window, within the
\SI{1000}{ms} budget), driving a live browser stream with no observed frame underruns over the tested run. A hybrid CPU--GPU deployment further runs the encoder on the CPU at the same \SI{24}{fps} and holds the decoder's one-shot CLIP/T5 conditioners on the CPU as well, cutting the per-GPU memory peak from \SI{21.1}{GB} to $\sim$\SI{7.7}{GB} and leaving the GPU for the diffusion critical path. Real-time decoding has been described as ``a common challenge for
generative codecs''~\cite{wang2025tgvc}; ReGenVC is, to our knowledge, the
first end-to-end generative video codec to overcome this challenge---combining
ultra-low-bitrate encoding with \emph{real-time} decoding on an 8-GPU node
built from consumer-class GPUs.
\end{abstract}

%% file: sec/1_intro.tex
\section{Introduction}
\label{sec:intro}

Video now dominates internet traffic, and the demand for ever-lower bitrates at
usable quality is relentless. Traditional hybrid codecs such as x264 and x265 are extremely mature, but they are fundamentally \emph{pixel-fidelity}
systems: at ultra-low bitrates their block-transform core degrades gracelessly
into blocking and ringing artifacts. For a talking-head clip, driving x264/x265
down to a few tens of kilobytes yields visibly broken video
(Sec.~\ref{sec:compress}). This is precisely the regime that matters for
bandwidth-starved settings---live streaming on weak links, video messaging,
telepresence---and precisely where traditional coding fails.

\emph{Generative video coding} offers a way out. Instead of transmitting pixel
residuals, the encoder sends a compact set of control signals and a strong
generative model at the receiver \emph{synthesizes} the video from
them~\cite{yi2025conditional,wu2026jscgc}. Because the heavy lifting is moved
from the bitstream to a learned prior, perceptual quality can be preserved at
bitrates unreachable by residual coding. The idea is compelling---but it has a
crippling practical problem: \textbf{the generative decoder is far too slow}. A
diffusion decoder must run a multi-step sampler over a large transformer and a
VAE for every group of frames, taking seconds per short clip on a single GPU.
For interactive use this is disqualifying, and it is the main reason generative
coding has remained a largely offline, conceptual proposition---a real-time
barrier that its own practitioners describe as ``a common challenge for
generative codecs''~\cite{wang2025tgvc}.

\paragraph{ReGenVC.}
We present \textbf{ReGenVC}, an end-to-end generative video codec for
talking-head video that targets the (bitrate\,$\times$\,latency) quadrant
left open by the generative codecs surveyed in Sec.~\ref{sec:related}
(Table~\ref{tab:landscape}): \emph{simultaneously} ultra-low-bitrate and
\emph{real-time-decodable}. On the
encoder side (Sec.~\ref{sec:encoder}) a source clip is reduced to three tiny
streams: a first frame compressed with a neural image codec,
per-frame body/face pose keypoints, and metadata---about \SI{26}{kB} for a 77-frame sequence, roughly $\tfrac{1}{10}$
of what x264/x265 need for essentially artifact-free reconstruction on the same clip. Overall the encoder side is comparatively lightweight and keeps up at
\SI{24}{fps} on 32 CPU cores alone, with no GPU involvement. On the decoder side a four-step
distilled Wan2.1-Fun-Control model~\cite{wan2025} reconstructs the video conditioned
on the transmitted pose and reference frame. The scientific and engineering core of
this paper is making that decoder run at \SI{24}{fps}.

\paragraph{Making the decoder real-time.}
Three coupled bottlenecks stand between a diffusion decoder and real time:
sampling depth (function evaluations multiply transformer cost), single-device
compute (one window of diffusion + VAE is too heavy for one GPU), and stage
serialization (conditioning encode, denoising, and pixel decode form a
dependency chain). We attack all three: \emph{(i)}~a
\emph{four-step distilled} transformer---obtained from the 20-step teacher via
our DMD2 + $x_0$-regression recipe (Sec.~\ref{sec:distill})---collapses the
sampler; techniques \emph{(ii)}--\emph{(iv)} then keep this 4-step student's
parameters unchanged: \emph{(ii)}~eight-GPU \emph{unified sequence parallelism}
(Ulysses$\times$Ring $=4\times2$)~\cite{jacobs2023ulysses,liu2023ringattention}
shards each forward; \emph{(iii)}~a \emph{spatially-split VAE} parallelizes
decode and encode; and \emph{(iv)}~a \emph{three-stage overlapped pipeline}
hides conditioning and cross-window encode behind the diffusion critical
path. This contrasts with
single-GPU real-time diffusion systems~\cite{kodaira2025streamdit,zhao2026sanastreaming,zhuang2025flashvsr},
which reach real time by \emph{changing the model} (quantization, sparsity,
architecture co-design) and thus trade away quality; we instead reach real
time on our 4-step student via \emph{model-preserving multi-GPU parallelism}. Orthogonal to per-window latency, we further hold the T5/CLIP one-shot conditioner weights on the CPU to free GPU memory for the diffusion critical path (Sec.~\ref{sec:hetero}).

\paragraph{Timing model.}
To reason about \emph{why} the pipeline does or does not fit the frame budget, we
contribute an analytical per-window timing model that decomposes wall-clock time
into diffusion, decode, encode, and their launch/overhead terms, yielding
$\wall \approx (w_d{+}d) + \max(v{+}a,\, w_e{+}e)$ (symbols defined in
Sec.~\ref{sec:timing}). It makes the real-time constraint $\wall \le \real$
analyzable and predicts the effect of each optimization before it is built.

\paragraph{Scope.}
We deliberately target \emph{talking-head / human} video, where pose is a strong
and compact driving signal; ReGenVC is not a general-purpose codec. Within this
scope it demonstrates that generative coding can be simultaneously
ultra-low-bitrate and real-time, which we believe is a necessary step toward
practical generative compression.

\paragraph{Contributions.}
\begin{itemize}
  \itemsep0.15em
  \item \textbf{An end-to-end generative video coding scheme}
  that transmits $\sim$\SI{26}{kB} for a 77-frame clip---about $10\times$ smaller
  than the $\sim$250--280\,kB x264/x265 need for artifact-free reconstruction,
  and sharp where they collapse at matched ultra-low bitrate
  (Sec.~\ref{sec:compress}).
  \item \textbf{A DMD2 + $x_0$-regression distillation recipe} that shrinks
  Wan2.1-Fun-Control's 20-step DDIM teacher into a 4-step student while
  mitigating the color-drift failure mode of vanilla DMD2 (Sec.~\ref{sec:distill}).
  \item \textbf{A real-time multi-GPU diffusion decoder} that sustains
  \SI{24}{fps} output on an 8-GPU node (\SI{972}{ms} per 25-frame window,
  within the \SI{1000}{ms} budget), by combining four-step distillation,
  eight-way USP, a spatially-split VAE, and a three-stage overlapped pipeline.
  \item \textbf{An analytical timing model} that decomposes per-window
  wall-clock time and characterizes the real-time feasibility region, validated
  against measured traces.
  \item \textbf{A hybrid CPU--GPU deployment}: on the encoder side, DWPose $+$
  \texttt{cheng2020\_attn} run entirely on CPU, with the streaming-critical DWPose
  stage sustaining \SI{24}{fps} on 32 CPU cores
  ($\approx$\SI{909}{ms} per 24 frames, within the \SI{1000}{ms} streaming
  budget); on the decoder side, the one-shot CLIP/T5 conditioner weights are
  held on the CPU as well, removing $\sim$\SI{13}{GB} of weights and
  cross-attention projections from GPU memory and cutting the per-GPU peak
  from \SI{21.1}{GB} to $\sim$\SI{7.7}{GB}.
\end{itemize}

%% file: sec/2_related.tex
\section{Related Work}
\label{sec:related}

\subsection{Generative video coding}
A mature parallel line, \emph{generative face video coding} (GFVC)~\cite{chen2024gfvc_review},
builds on motion-keypoint driven animation techniques
(FOMM~\cite{siarohin2019fomm}, Face-vid2vid~\cite{wang2021face_vid2vid}) and is
now under active standardization in JVET. Our work is complementary: we
target the same talking-head domain and ultra-low-bitrate regime, but replace
the low-capacity keypoint-warping decoder with a distilled video-diffusion
transformer---a heavier but stronger generative prior---while still meeting a
real-time budget through multi-GPU parallelism.

Reframing compression as conditional generation more broadly is an active and fast-moving
frontier. Recent systems drive a generative prior from compact conditions:
Yi \etal~\cite{yi2025conditional} synthesize video from compact conditions with a \SI{14}{B} diffusion backbone;
JSCGC~\cite{wu2026jscgc} unifies source, channel, and generation coding;
M3-CVC~\cite{wan2024m3cvc} steers a text-conditioned diffusion decoder with a
multimodal LLM; DiSCo~\cite{wang2025disco} decomposes a clip into text, a
degraded video, and optional pose/sketch cues for a conditional video-diffusion
decoder; T-GVC~\cite{wang2025tgvc} guides a diffusion decoder with
sparse motion trajectories; and training-free or zero-shot schemes such as
Free-GVC~\cite{ling2026freegvc} and ZeroGVC~\cite{gao2026zerogvc} compress along
a video-diffusion latent trajectory. These sit alongside strong neural codecs
such as DCVC-FM~\cite{li2024dcvcfm}, while traditional hybrid codecs (x264/x265)
remain the deployed baseline but degrade into blocking at ultra-low bitrate.
Two properties characterize this literature. First, it optimizes
\emph{rate-distortion-perception} on \emph{offline} reconstructions and almost
never reports decoder \emph{latency}: none of these systems runs in real
time. Second, most target general
video with heavy or general-purpose priors (multimodal LLMs, multi-billion
parameter backbones, per-clip trajectory optimization), and several works cast
domain specialization as a limitation to overcome. We take the opposite stance:
we deliberately specialize a \emph{compact} pose-controlled backbone to the
talking-head domain, trading that generality for the parallelism headroom that
makes real-time decoding possible. We therefore differ by delivering a
\emph{complete} codec whose generative decoder runs in \emph{real time}
on a single 8-GPU node---filling the latency gap these systems leave open---rather
than by claiming state-of-the-art offline reconstruction quality; we
quantify the gap to x264/x265 in Sec.~\ref{sec:compress}.

\begin{table*}[t]
  \centering\small
  \setlength{\tabcolsep}{4pt}
  \renewcommand{\arraystretch}{1.15}
  \caption{Where ReGenVC sits in the landscape of talking-head/video compression and streaming diffusion.
  ``Ultra-low bitrate'' means the method operates at $\lesssim$\SI{50}{kB} for a $\sim$3\,s clip \emph{with recognizable output} (a traditional codec that produces blocking artifacts at that rate is marked \ding{55} here even though the bitstream is small); ``real-time'' means sustained decoding at video frame rate.
  Only ReGenVC (last row) is a \emph{generative video codec} that is simultaneously ultra-low-bitrate and real-time-decodable;
  the keypoint-warping line is both but has no diffusion prior, and every surveyed diffusion-based generative codec reports offline decoding.
  Symbols: \ding{51} property holds, \ding{55} does not, --- not applicable.}
  \label{tab:landscape}
  \begin{tabular}{@{}p{5.8cm}p{3.4cm}ccc@{}}
    \toprule
    Method & Category & Ultra-low bitrate & Real-time & Diffusion prior \\
    \midrule
    x264 / x265 (FFmpeg) & traditional codec & \ding{55}\rlap{$^{a}$} & \ding{51} & --- \\
    FOMM~\cite{siarohin2019fomm}, Face-vid2vid~\cite{wang2021face_vid2vid} (GFVC line) & gen.\ face codec (warping) & \ding{51} & \ding{51}\rlap{$^{c}$} & \ding{55} \\
    Yi'25~\cite{yi2025conditional}, M3-CVC~\cite{wan2024m3cvc}, DiSCo~\cite{wang2025disco}, T-GVC~\cite{wang2025tgvc}, Free-GVC~\cite{ling2026freegvc}, ZeroGVC~\cite{gao2026zerogvc} & gen.\ video codec (diffusion) & \ding{51} & \ding{55}\rlap{$^{b}$} & \ding{51} \\
    StreamDiT~\cite{kodaira2025streamdit}, SANA-Streaming~\cite{zhao2026sanastreaming}, FlashVSR~\cite{zhuang2025flashvsr} & streaming diffusion (not a codec) & --- & \ding{51} & \ding{51} \\
    \midrule
    \textbf{ReGenVC (ours)} & \textbf{gen.\ video codec (diffusion)} & \textbf{\ding{51}} & \textbf{\ding{51}} & \textbf{\ding{51}} \\
    \bottomrule
  \end{tabular}\\[2pt]
  {\footnotesize $^{a}$ x264/x265 remain real-time and can produce the same $\sim$\SI{26}{kB} but collapse into blocking there (Sec.~\ref{sec:compress}).
  $^{b}$ Reported decoder latencies from the cited papers, all below video frame rate: Yi'25~\cite{yi2025conditional} states its ``decoding speed falls short of real-time requirements''; M3-CVC~\cite{wan2024m3cvc} reports \SI{142.5}{s} decoder latency per clip; DiSCo~\cite{wang2025disco} reports \SI{396.20}{ms} per frame ($\sim$2.5\,fps); T-GVC~\cite{wang2025tgvc} calls real-time decoding ``a common challenge for generative codecs''; Free-GVC~\cite{ling2026freegvc} reports 1.25\,fps and notes that ``real-time processing remains challenging''; ZeroGVC~\cite{gao2026zerogvc} reports \SI{202}{ms} per frame at $N{=}6$ ($\sim$5\,fps). JSCGC~\cite{wu2026jscgc} demonstrates on image transmission and does not report a video decoding rate, so we discuss it in the text but omit it from this row.
  $^{c}$ Real-time for the GFVC line follows from the compact keypoint-plus-warping architecture and Face-vid2vid's explicit video-conferencing target (``better suited for video streaming'', matching H.264 quality at one-tenth of the bandwidth); neither paper reports an explicit inference frame rate.}
\end{table*}

\subsection{Diffusion distillation and few-step sampling}
The dominant cost of diffusion inference~\cite{ho2020ddpm,rombach2022ldm,lipman2023flowmatching} is the number of function evaluations.
A large body of work compresses the sampling trajectory. Progressive
distillation~\cite{salimans2022progressive} halves the step count repeatedly;
consistency models~\cite{song2023consistency} and their latent
variant~\cite{luo2023lcm} learn a direct map from any point on the trajectory to
its endpoint, enabling one- or few-step generation. Distribution-matching
distillation (DMD/DMD2)~\cite{yin2024dmd,yin2024dmd2} trains a few-step student
by matching the score of the teacher's output distribution, and
InstaFlow~\cite{liu2023instaflow} straightens the probability-flow ODE via
reflow. In the video regime, CausVid~\cite{yin2025causvid} and
Self-Forcing~\cite{huang2025selfforcing} convert bidirectional video diffusion
into fast autoregressive students. We build on DMD2 for two reasons: it reaches the few-step regime in a single
training run (rather than by iterated halving as in progressive distillation),
and it leaves the student's backbone architecture identical to the teacher's. We adapt it to our control-video
setting, augmenting it with an $x_0$-latent regression loss that anchors the
student's $x_0$ prediction to the teacher's and mitigates a color-drift failure
mode we observed with vanilla DMD2. This 4-step distilled student is the sampler on which our systems
techniques operate; details are given in Sec.~\ref{sec:distill}.

\subsection{Real-time and streaming video diffusion}
Serving diffusion interactively has driven a family of pipeline designs.
StreamDiffusion~\cite{kodaira2024streamdiffusion} introduced stream batching and
pre-computation for real-time interactive image generation.
StreamDiT~\cite{kodaira2025streamdit} generates streaming text-to-video with a
moving buffer and reaches \SI{16}{fps};
SANA-Streaming~\cite{zhao2026sanastreaming} co-designs a hybrid diffusion
transformer that interleaves softmax and linear attention blocks with
mixed-precision (FP4/FP8) quantization for \SI{24}{fps} video editing;
FlashVSR~\cite{zhuang2025flashvsr} delivers one-step streaming super-resolution.
These systems reach real time by aggressively \emph{modifying the model}
---quantization, sparsity, or heavy compression---which trades away fidelity,
and they target open-ended generation or editing. Our approach differs in two
respects: our target is a \emph{conditional generative decoder} (compression
backend); and while we too distill the teacher to a 4-step student
(Sec.~\ref{sec:distill}), we then reach real time on that student with
\emph{model-preserving multi-GPU parallelism}---rather than further shrinking it via
quantization or sparsity---which raises the cross-device scheduling and
communication issues we address (Sec.~\ref{sec:overlap}).

\subsection{Parallel and efficient diffusion inference}
Scaling diffusion transformers across devices builds on sequence-parallel
attention. DeepSpeed-Ulysses~\cite{jacobs2023ulysses} shards the sequence with
all-to-all communication, while Ring attention~\cite{liu2023ringattention}
streams key/value blocks around a ring; unified sequence parallelism composes
the two. xDiT~\cite{fang2024xdit} and PipeFusion~\cite{wang2024pipefusion} bring
these ideas to DiT inference with patch-level pipelining and reuse of stale
feature maps across diffusion steps. Efficient attention kernels~\cite{dao2022flashattention,dao2023flashattention2}
underpin all of the above. We adopt USP for the transformer, but extend the
parallel scope to the \emph{VAE} via spatial splitting and to
\emph{cross-window} scheduling of the decode and next-window encode.

%% file: sec/3_method.tex
\section{Method}
\label{sec:method}

\begin{figure*}[t]
  \centering
  \begin{tikzpicture}[
      font=\small,
      box/.style={draw,rounded corners,minimum height=8mm,minimum width=20mm,align=center},
      enc/.style={box,fill=orange!14},
      bit/.style={box,fill=gray!14,minimum width=26mm},
      diff/.style={box,fill=blue!12},
      vae/.style={box,fill=green!14},
      arrow/.style={-{Latex[length=1.6mm]},thick},
    ]
    \node[box,fill=black!5] (src) {source\\video};
    \node[enc] (ff) [right=6mm of src,yshift=6mm] {first frame\\$\to$\SI{10}{kB}};
    \node[enc] (pose) [right=6mm of src,yshift=-6mm] {pose keypoints\\$\to$\SI{15.8}{kB}};
    \node[bit] (bs) [right=31mm of src] {\textbf{bitstream}\\$\approx$\SI{26}{kB}};
    \node[diff] (dit) [right=8mm of bs] {4-step distilled\\DiT (USP $4{\times}2$)};
    \node[vae] (dec) [right=7mm of dit] {split VAE\\decode};
    \node[box,fill=black!5] (out) [right=7mm of dec] {output\\video};
    \draw[arrow] (src) -- (ff); \draw[arrow] (src) -- (pose);
    \draw[arrow] (ff) -- (bs); \draw[arrow] (pose) -- (bs);
    \draw[arrow] (bs) -- (dit); \draw[arrow] (dit) -- (dec); \draw[arrow] (dec) -- (out);
    \node[draw,dashed,rounded corners,fit=(ff)(pose),inner sep=2mm,label=above:{\footnotesize Encoder (CPU, streaming)}] {};
    \node[draw,dashed,rounded corners,fit=(dit)(dec),inner sep=2mm,label=above:{\footnotesize Decoder (real-time, $8$ GPUs)}] {};
  \end{tikzpicture}
  \caption{ReGenVC end-to-end codec. The encoder transmits only identity (a neurally
  compressed reference frame), motion (pose keypoints), and a small metadata header,
  totalling $\approx$\SI{26}{kB} for a 77-frame clip. The decoder regenerates the
  video with a four-step distilled diffusion transformer and a spatially-split VAE,
  in real time on $8$ GPUs.}
  \label{fig:codec}
\end{figure*}
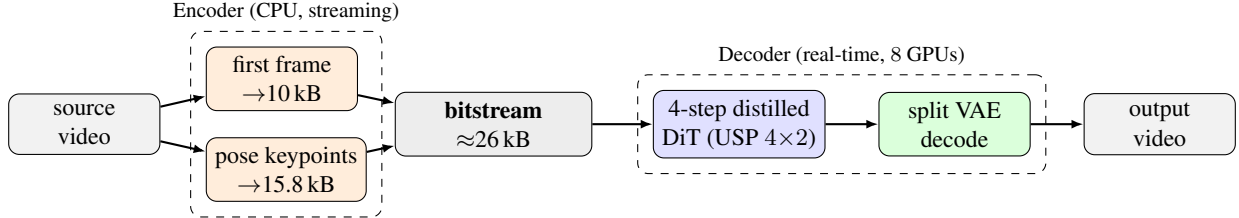

\subsection{End-to-end codec overview}
\label{sec:codec}
ReGenVC is a complete codec: an encoder maps a source clip to a compact bitstream and
a decoder reconstructs the clip from it (Fig.~\ref{fig:codec}). The design
principle is to transmit only what a strong generative prior \emph{cannot} infer
---identity/appearance (a single reference frame) and motion (pose keypoints)---and
to regenerate everything else (texture, temporal coherence, view-consistent
detail) at the receiver. On the decode side this reference frame is held \emph{fixed}: encoded once (via CLIP's image encoder; Sec.~\ref{sec:encoder}) and used as visual conditioning at every window, it keeps the whole stream tied to one identity and appearance, while its one-time cost stays off the per-window critical path. This is what enables the order-of-magnitude bitrate
reduction over pixel-fidelity coding, at the cost of restricting the domain to
pose-drivable (talking-head/human) video.

\subsection{Encoder}
\label{sec:encoder}
The encoder produces three streams (sizes in Sec.~\ref{sec:compress}):
\emph{(i)}~a \textbf{reference frame}---the first frame compressed with the neural
image codec \texttt{cheng2020\_attn}~\cite{cheng2020} (CompressAI~\cite{begaint2020compressai}) at quality~6
($\sim$\SI{10}{kB}), carrying identity and appearance; \emph{(ii)}~\textbf{pose}
---per-frame body/face keypoints from DWPose~\cite{yang2023dwpose} serialized as
quantized coordinates ($\sim$\SI{15.8}{kB} for 77 frames), replacing a full pose
\emph{video} with a few hundred bytes per frame and carrying motion; and
\emph{(iii)}~\textbf{metadata} ($\sim$\SI{0.2}{kB}). The total is
$\sim$\SI{26}{kB}. The encoder runs entirely on the CPU (Sec.~\ref{sec:hetero}); pose extraction dominates its cost. At the receiver
the reference frame is decoded once and the pose stream is rendered to conditioning
canvases that drive the diffusion decoder (Sec.~\ref{sec:setup}--\ref{sec:compile}). How these signals condition the diffusion model---pose through a ControlNet-style control branch~\cite{zhang2023controlnet}, the reference frame through CLIP's image encoder~\cite{radford2021clip}, and any text context through CLIP's text branch and T5~\cite{raffel2020t5}---follows the Wan2.1-Fun-Control model~\cite{wan2025} and is not a contribution of this work.

\subsection{Problem setup and the real-time budget}
\label{sec:setup}
We serve a conditional video-diffusion decoder in a \emph{sliding-window}
regime. The stream is processed in windows of $F=25$ latent-aligned frames (i.e., frame counts and indices coincide with the video-VAE's latent grid).
For each window the system (i)~prepares conditioning signals (pose/control
encode), (ii)~runs a distilled diffusion sampler over a transformer backbone to
denoise the window latent, and (iii)~decodes the latent to pixels through a VAE.
To keep the generation self-consistent across windows, the decoder also encodes
an \emph{anchor} region of the freshly generated frames back into latent space
to condition the next window; consecutive windows overlap by this single anchor
frame, so each window advances the stream by $S = F - 1 = 24$ new frames. Two VAE encodes therefore cross the window boundary and feed the next diffusion pass: an \textbf{anchor encode} of the freshly generated overlap frames (which seed the next window for temporal continuity), and a \textbf{conditioning encode} of the next window's pose/control canvases.

At a target frame rate $\text{fps}=24$, a window's $S$ net frames must be produced
within their playback duration, giving the per-window real-time budget
\begin{equation}
  \real \;=\; \frac{S}{\text{fps}} \;=\; \frac{F-1}{\text{fps}}
        \;=\; \frac{24}{24}\,\text{s}
        \;=\; \SI{1000}{ms}.
  \label{eq:budget}
\end{equation}
Real-time serving requires the \emph{steady-state} per-window wall-clock time
\wall{} (whose steady-state decomposition we derive in Sec.~\ref{sec:timing}) to satisfy $\wall \le \real$ for every window, so that the output frame
queue never underruns. This is a throughput constraint on a producer feeding a
fixed-rate consumer; transient windows may exceed \real{} only if absorbed by a
bounded jitter buffer. Our objective is therefore to \emph{minimize and stabilize}
\wall.

\subsection{Four-step distilled diffusion transformer}
\label{sec:distill}
The backbone is the transformer of a pretrained control model~\cite{wan2025}, a video diffusion transformer that follows the evolution of latent video diffusion from U-Net designs like SVD~\cite{blattmann2023svd} to transformer backbones like CogVideoX~\cite{yang2024cogvideox}.
A naive sampler evaluates this transformer for tens of steps per window, which
alone exceeds \real. We instead distill the 20-step DDIM teacher into a \emph{four-step
student} that reproduces the teacher's sampling trajectory with a fixed
four-point timestep schedule $\{t_1,t_2,t_3,t_4\}$. Our recipe applies
distribution-matching distillation (DMD2)~\cite{yin2024dmd2} augmented with an
$x_0$-latent regression loss:
\begin{equation}
\mathcal{L}_\text{gen} = \mathcal{L}_\text{DMD2} + \lambda \,
\mathcal{L}_{\text{reg}}(x_0^\text{stu},\, x_0^\text{tea}).
\end{equation}
We found that vanilla DMD2 in this control-video setting produces visible
\emph{color drift} across denoising trajectories; the regression term
anchors the student's $x_0$ prediction to the teacher's, breaks the DMD
adversarial equilibrium, and stabilizes color. Crucially for a serving system,
the student retains the teacher's \emph{backbone architecture}, so all
downstream parallelization and compilation apply identically to it. Reducing the
number of function evaluations from $\sim\!20$ to $4$ is the single largest
latency reduction in the system and is the enabling precondition for real-time
operation; the remaining sections make those four evaluations, and the VAE that
surrounds them, fit the budget.

\paragraph{Distillation setup (reproducibility).}
The student samples at four fixed timesteps
$\{t_1,t_2,t_3,t_4\}=\{1000, 750, 500, 250\}$ selected from the teacher's
20-step DDIM schedule.
We train the student with the objective above (regression weight $\lambda=10.0$)
at learning rate $1\!\times\!10^{-5}$ with 100-step linear warmup, in \texttt{bf16},
using per-device batch size $1$ with gradient accumulation $4$ (effective batch
$4$). Training runs on a single GPU for 500 steps and we deploy the checkpoint
at step 300 (selected on held-out reconstructions). The distillation corpus is
$\sim$7{,}181 short person-motion video/control pairs (mp4) derived from
publicly-available human-motion clips; the resulting student is applied at test
time to the talking-head domain (Sec.~\ref{sec:compress}), which is a mild
in-modality generalization from the training distribution.
Distillation is a one-time offline cost and does not affect the served
pipeline described below.

\subsection{Multi-GPU unified sequence parallelism}
\label{sec:usp}
Running the four-step sampler on one GPU exceeds the budget. We shard the \emph{token sequence} of the diffusion
transformer across $N=8$ GPUs using unified sequence parallelism (USP), which
composes two orthogonal mechanisms:
\begin{itemize}
  \itemsep0.1em
  \item \textbf{Ulysses}~\cite{jacobs2023ulysses}: an all-to-all redistributes
  the sequence dimension into the head dimension before attention and back
  afterwards, so each GPU computes exact attention over the full sequence for a
  subset of heads. We use an Ulysses degree $U=4$.
  \item \textbf{Ring attention}~\cite{liu2023ringattention}: the remaining
  factor streams key/value blocks around a ring of GPUs, computing attention in
  an online-softmax fashion. We use a ring degree $R=2$.
\end{itemize}
The product $U\times R = 4\times2 = 8$ tiles the sequence across all eight
devices. This hybrid is preferred over pure Ulysses (whose degree is capped by
the head count and whose all-to-all volume grows with device count) and pure
Ring (whose latency is sensitive to the slowest hop). We adopt the $4\times2$
configuration on our node: on our backbone $U{=}4$ stays below the head-count
cap, and we prefer larger $U$ over larger $R$ to keep the Ring's per-hop
latency low on our 8-GPU NVLink topology. Because USP preserves \emph{exact} attention, the parallel
student is numerically equivalent to the single-device student up to reduction
order.

\subsection{Spatially-split VAE decode and encode}
\label{sec:vae}
After denoising, the window latent must be decoded to pixels by the VAE, and an
anchor region must be encoded back to latent to condition the next window. On a
single GPU these VAE passes are surprisingly expensive---comparable to several
transformer forwards (Sec.~\ref{sec:exp})---because the decoder operates at full
spatial resolution. We therefore \emph{spatially split} the VAE: the latent (or
image) is partitioned along the spatial dimensions into $N$ tiles, each GPU
decodes/encodes its tile, and results are gathered. To avoid seams at tile
boundaries we include a halo overlap sized to the VAE receptive field and blend
the overlap on gather. Spatial splitting turns the VAE from a serial per-device
cost into a parallel one, and---unlike sequence splitting---requires
communication only at the split and gather points rather than inside every
attention layer.

\subsection{Three-stage overlapped pipeline and cross-communicator scheduling}
\label{sec:overlap}
\begin{figure*}[t]
  \centering
  \begin{tikzpicture}[
    x=0.0055cm, y=0.55cm,
    font=\footnotesize,
  ]
    \fill[gray!5]  (0,-2.7) rectangle (1000, 0.9);
    \fill[gray!10] (1000,-2.7) rectangle (2000, 0.9);
    \draw[dashed,gray!70] (0,-2.7) -- (0,0.9);
    \draw[dashed,gray!70] (1000,-2.7) -- (1000,0.9);
    \draw[dashed,gray!70] (2000,-2.7) -- (2000,0.9);
    \draw[->,-{Latex[length=1.4mm]},thick] (-30,-3) -- (2100,-3);
    \node[right,font=\tiny] at (2100,-3) {time (ms)};
    \foreach \x in {0,500,1000,1500,2000}{
      \draw (\x,-2.95) -- (\x,-3.05);
      \node[below,font=\tiny] at (\x,-3.05) {\x};
    }
    \node[left,font=\tiny,align=right] at (-30, 0.3) {(iii) $\text{cpu}$};
    \node[left,font=\tiny,align=right] at (-30,-0.75) {(ii) $d{+}v{+}a$};
    \node[left,font=\tiny,align=right] at (-30,-2.0) {(i) $p$};
    \draw[fill=blue!25]  (0,-1.1) rectangle (513,-0.4);
    \node[font=\tiny] at (256,-0.75) {$w_d{+}d$};
    \draw[fill=green!30] (513,-1.1) rectangle (972,-0.4);
    \node[font=\tiny] at (744,-0.75) {$\max(v{+}a,\,w_e{+}e)$};
    \draw[fill=orange!30] (0,-2.3) rectangle (972,-1.7);
    \node[font=\tiny] at (488,-2.0) {$p$ for window $N{+}1$};
    \draw[fill=gray!45] (0,0.0) rectangle (972,0.6);
    \node[font=\tiny] at (488,0.3) {D2H copy $+$ numpy for window $N{-}1$};
    \draw[fill=blue!25]   (1000,-1.1) rectangle (1513,-0.4);
    \draw[fill=green!30]  (1513,-1.1) rectangle (1972,-0.4);
    \draw[fill=orange!30] (1000,-2.3) rectangle (1972,-1.7);
    \draw[fill=gray!45]   (1000,0.0) rectangle (1972,0.6);
    \draw[<->,thick] (0,1.15) -- (1000,1.15);
    \node[above,font=\tiny] at (500,1.15) {steady-state window ($\real=\SI{1000}{ms}$)};
  \end{tikzpicture}
  \caption{Steady-state three-stage pipeline (Sec.~\ref{sec:overlap}). In every window three different windows are in flight on independent resources: stage~(i) prepares the \emph{next} window's pose on the CPU (the next-window VAE encode $e$ appears inside stage~(ii)'s $\max$ path); stage~(ii) runs diffusion followed by the overlapped VAE decode/encode for the \emph{current} window on the main and \texttt{dec\_stream}; stage~(iii) copies-out and post-processes the \emph{previous} window on the decoder CPU thread. Only stage~(ii) is on the wall-clock critical path: $\wall \approx (w_d{+}d) + \max(v{+}a,\,w_e{+}e)$; segment widths reflect measured values from Table~\ref{tab:e2e}.}
  \label{fig:timeline}
\end{figure*}
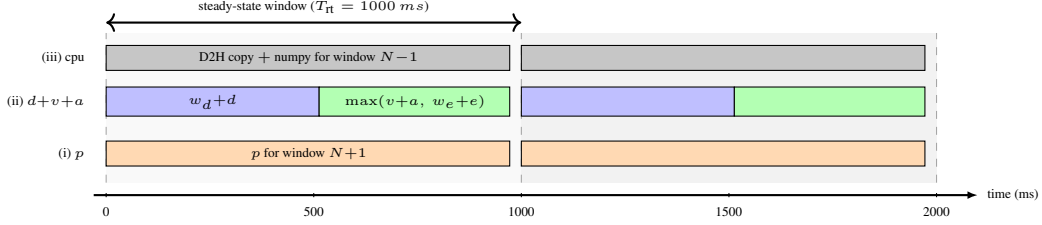

The three components of a window---conditioning encode, diffusion, and VAE
decode---form a dependency chain, and additionally the \emph{next} window's
anchor encode depends on the \emph{current} window's decoded pixels. Executed
serially, encode + diffusion + decode + anchor-encode exceeds \real. We break this
with a \emph{three-stage software pipeline}. The three stages run on
independent resources: \emph{(i)}~\textbf{pose preparation} ($p$),
executed by the encoder worker on the CPU;
\emph{(ii)}~\textbf{diffusion, decoding, and next-window anchor} ($d{+}v{+}a$),
executed by the main thread and decoder worker on the main CUDA stream and
\texttt{dec\_stream} (with $v$ and $a$ strictly following $d$); and
\emph{(iii)}~\textbf{host post-processing} ($\text{cpu}$), the D2H copy and
numpy conversion carried out on the decoder worker's CPU thread. In steady
state, the same wall-clock interval is occupied by three different windows at
these three stages---the next window's $p$, the current window's
$d{+}v{+}a$, and the previous window's $\text{cpu}$---all making progress on
independent resources. Within stage \emph{(ii)}, the current window's decode
branch ($v_k{+}a_{k+1}$) on \texttt{dec\_stream} runs concurrently with the
next window's encode ($e_{k+1}$) on \texttt{enc\_stream}, and each split-VAE
collective is spread across the eight GPUs as in Sec.~\ref{sec:vae}. The
critical path therefore pays only $\max(v{+}a,\, w_e{+}e)$
(see Sec.~\ref{sec:timing}), not their sum. Figure~\ref{fig:timeline}
visualizes one steady-state window of this three-stage overlap.

Overlapping two spatially-split VAE stages means \emph{two independent
collective operations} (the gather/scatter of decode and encode) are in flight
on \emph{different communicators} at the same time. Naively enqueuing them from
independent Python threads can deadlock: if rank~$0$ enters the decode collective
while rank~$1$ enters the encode collective first, the two communicators can
wait on each other. We avoid this with a \emph{deterministic global enqueue
order}: every rank enqueues the decode and encode collectives in the same fixed
order, guaranteeing progress while still allowing the underlying runtime to
execute them concurrently on separate streams.

\subsection{Graph compilation}
\label{sec:compile}
Finally, we compile the transformer with \texttt{torch.compile}. We find that
\texttt{mode=default} yields a robust reduction in per-window wall-clock time by
fusing pointwise operations and reducing Python/launch overhead, whereas
\texttt{mode=reduce-overhead} (CUDA-graph capture) is \emph{unusable} here: the
graph's static tensor lifetimes conflict with the pipeline's cross-stream tensor
reuse (notably in the sinusoidal timestep embedding), causing a correctness/lifetime
fault. We therefore standardize on \texttt{mode=default}, which is safe under our
multi-stream, cross-communicator schedule. The net effect is a
\SI{33}{ms} per-window reduction (Sec.~\ref{sec:exp}), turning a marginally
over-budget pipeline into a real-time one.

\subsection{Hybrid CPU--GPU deployment}
\label{sec:hetero}
ReGenVC targets a hybrid CPU\,$+$\,GPU node that keeps the GPU
dedicated to the repeated diffusion critical path and offloads everything else to
the CPU. Two classes of work move off the GPU. \textbf{(i)~The entire encoder:}
its models are small---CompressAI \texttt{cheng2020\_attn} (\SI{31.7}{M}) and DWPose
(YOLOX-L\,$+$\,DW-LL, \SI{87.8}{M})---and run efficiently on the CPU. Only the per-frame DWPose extractor lies on the streaming critical path; on 32 CPU cores it sustains \SI{24}{fps}, matching the streaming budget, while the one-shot \texttt{cheng2020\_attn} first-frame compression runs once per clip and is off the critical path. Encoding is therefore entirely GPU-free.
\textbf{(ii)~The one-shot conditioners on the decode side:} the CLIP
(ViT-H/14~\cite{radford2021clip,ilharco2021openclip,conneau2020xlmr}, \SI{1.2}{B}) and T5 (umt5-xxl encoder~\cite{raffel2020t5}, $\sim$\SI{5.3}{B}) text/image encoders are
evaluated \emph{once per sequence}, not per denoising step. We keep their weights
and computation on the CPU and stream only the resulting cross-attention
key/value tensors to the GPU (host-to-device)---\SI{94.4}{MB} for T5 and
\SI{47.4}{MB} for CLIP in \texttt{bf16}---instead of holding the
$\sim$$5.3\text{B}{+}1.2\text{B}$ conditioner weights in GPU memory. In our measurements this offload removes \SI{12.8}{GB} of conditioner weights (plus \SI{0.54}{GB} of key/value projections) from the GPU, cutting the per-GPU memory peak from \SI{21.1}{GB} to $\sim$\SI{7.7}{GB} (Table~\ref{tab:mem}). The GPU
footprint is thus dominated by the four-step transformer and the split VAE rather
than by conditioning models. In contrast to single-GPU real-time diffusion systems~\cite{kodaira2025streamdit,zhao2026sanastreaming,zhuang2025flashvsr}, which keep all conditioning models resident in GPU memory or shrink them via quantization, we hold the T5/CLIP conditioner weights on the CPU---freeing GPU memory without lossily compressing them.

\begin{table}[t]
  \centering\small
  \setlength{\tabcolsep}{4pt}
  \caption{Per-GPU memory footprint and the effect of CPU offloading (measured).
  Relocating the one-shot T5 and CLIP conditioners to the CPU removes their
  weights from the GPU, cutting the peak from \SI{21.1}{GB} to $\sim$\SI{7.7}{GB}
  and leaving the budget to the four-step transformer and the VAE.}
  \label{tab:mem}
  \begin{tabular}{@{}lcc@{}}
    \toprule
    component & GPU memory & placement \\
    \midrule
    Transformer (DiT 1.3B\,$+$\,Ctrl) & \SI{2.91}{GB} & GPU \\
    VAE (Wan2.1 enc.\,$+$\,dec.) & \SI{0.24}{GB} & GPU \\
    T5 (umt5-xxl encoder) & \SI{10.58}{GB} & $\to$ CPU \\
    CLIP (ViT-H/14\,$+$\,XLM-R) & \SI{2.22}{GB} & $\to$ CPU \\
    \midrule
    steady-state weights & \SI{15.95}{GB} & \\
    absolute peak & \SI{21.07}{GB} & \\
    \textbf{after CPU offload} & \boldmath$\sim$\textbf{\SI{7.7}{GB}} & \\
    \bottomrule
  \end{tabular}
\end{table}

%% file: sec/4_timing.tex
\section{Analytical Timing Model}
\label{sec:timing}

To reason about real-time feasibility we decompose the per-window wall-clock
time into GPU-compute terms and their associated host-side overhead/wait terms.
This model lets us (i)~identify the critical-path stage, (ii)~predict the effect
of an optimization before implementing it, and (iii)~validate the pipeline by
checking that measured windows obey it.

\subsection{Terms}
For one window we define the GPU-time terms
\begin{center}
\small
\begin{tabular}{@{}ll@{}}
\toprule
symbol & meaning (GPU compute time) \\
\midrule
$d$ & diffusion: four transformer forwards under USP \\
$v$ & VAE decode of the current window (spatially split) \\
$a$ & anchor encode of the generated frames \\
$e$ & VAE encode of the \emph{next} window's condition (split) \\
\bottomrule
\end{tabular}
\end{center}
and the host-side overhead/wait terms
\begin{center}
\small
\begin{tabular}{@{}ll@{}}
\toprule
symbol & meaning (launch / wait overhead) \\
\midrule
$p$ & pose/control preparation on the CPU (rendering + H2D upload) \\
$w_d$ & Python + kernel-launch overhead around diffusion \\
$w_e$ & extra wait on the encode stream beyond $e$ \\
$\text{cpu}$ & device-to-host copy + numpy post-processing \\
\bottomrule
\end{tabular}
\end{center}
Here $w_e = \text{wait\_enc} - e$ captures the portion of the encode wait
not explained by encode compute (cross-stream serialization and synchronization
slack); $w_d$ analogously accounts for Python and kernel-launch overhead
surrounding diffusion on the main stream, and is obtained as the residual
required by Eq.~\eqref{eq:wall} given the other measured terms.

\subsection{Wall-clock decomposition}
Because decode ($v$) and next-window encode ($e{+}w_e$) are placed on separate
streams and overlapped (Sec.~\ref{sec:overlap}), while diffusion ($d$) with its
launch overhead ($w_d$) is on the critical path preceding them, the steady-state
per-window time is
\begin{equation}
  \wall \;\approx\; \underbrace{(w_d + d)}_{\text{diffusion path}}
        \;+\; \underbrace{\max\!\big(v + a,\; w_e + e\big)}_{\text{overlapped VAE path}}
  \label{eq:wall}
\end{equation}
The $\max$ captures the overlap: the window pays for the \emph{longer} of the
decode branch ($v{+}a$) and the concurrent next-window encode branch
($w_e{+}e$), not their sum. The conditioning preparation $p$ and the host
post-processing $\text{cpu}$ are either folded into $w_d$ or absorbed by the
three-stage pipeline's concurrent processing of adjacent windows
(Sec.~\ref{sec:overlap}), and thus do not appear on the steady-state
critical path.

\subsection{Real-time feasibility}
Combining Eq.~\eqref{eq:budget} and Eq.~\eqref{eq:wall}, real-time serving
requires
\begin{equation}
  (w_d + d) + \max(v + a,\; w_e + e) \;\le\; \real = \SI{1000}{ms}.
  \label{eq:feasible}
\end{equation}
Equation~\eqref{eq:feasible} exposes the design levers directly:
\begin{itemize}
  \itemsep0.1em
  \item \textbf{Distillation} shrinks $d$ (fewer NFEs) and is the dominant term
  in the diffusion path.
  \item \textbf{USP width} shrinks $d$ further by sharding each forward.
  \item \textbf{VAE spatial split} shrinks $v$ and $e$, lowering both arguments
  of the $\max$.
  \item \textbf{Overlap} replaces a serial cost $v{+}a{+}w_e{+}e$ with $\max(v{+}a,\, w_e{+}e)$; if the branches are balanced this nearly halves the VAE contribution.
  \item \textbf{Compilation} shrinks $w_d$ by cutting Python/launch overhead.
\end{itemize}

%% file: sec/5_experiments.tex
\section{Experiments}
\label{sec:exp}

\subsection{Setup}
Our serving node has $8\times$ NVIDIA RTX~5090D GPUs (Blackwell, \texttt{sm\_120},
\SI{32}{GB} each) paired with an Intel Xeon~6 CPU (32-core, AMX). The backbone is a distilled four-step control model derived
from Wan2.1-Fun-Control~\cite{wan2025}. All inference uses \texttt{bfloat16}. Unless noted,
the transformer runs under $8$-way USP ($U{=}4$, $R{=}2$), the VAE decode and
encode are spatially split across all $8$ GPUs, decode/encode of neighboring
windows are overlapped on separate streams, and the transformer is compiled with
\texttt{torch.compile(mode=default)}. Windows contain $F=25$ frames, giving the
real-time budget $\real=\SI{1000}{ms}$ (Eq.~\eqref{eq:budget}). We report
steady-state windows (after pipeline warm-up) to characterize sustained
throughput.

\paragraph{Test clips.}
We use two fixed-camera single-subject talking-head interview
clips. The quantitative compression comparison (Sec.~\ref{sec:compress}) is
conducted on a 77-frame ($\sim$3\,s at 24\,fps) clip, chosen because it has a
fixed camera, no scene cuts, and a single subject---the regime for which
ReGenVC's pose-drivable design is intended. For the live serving demo
(Fig.~\ref{fig:demo}) and the temporal reconstruction figure
(Fig.~\ref{fig:recon}) we use a longer 240-frame ($\sim$10\,s at 24\,fps) clip
of the same type to visualize sustained decoding over multiple windows; its
bitstream is $\sim$\SI{60}{kB} ($\sim$\SI{10}{kB} reference + $\sim$\SI{49}{kB}
pose + $\sim$\SI{0.2}{kB} meta), the same per-frame pose cost and one-shot
reference-frame cost as the 77-frame clip. A benchmark-scale evaluation across
a talking-head dataset is out of scope for this paper and is left as future work.

\subsection{Compression: bitrate and quality vs.\ x264/x265}
\label{sec:compress}
\begin{figure*}[t]
  \centering
  \includegraphics[width=\linewidth]{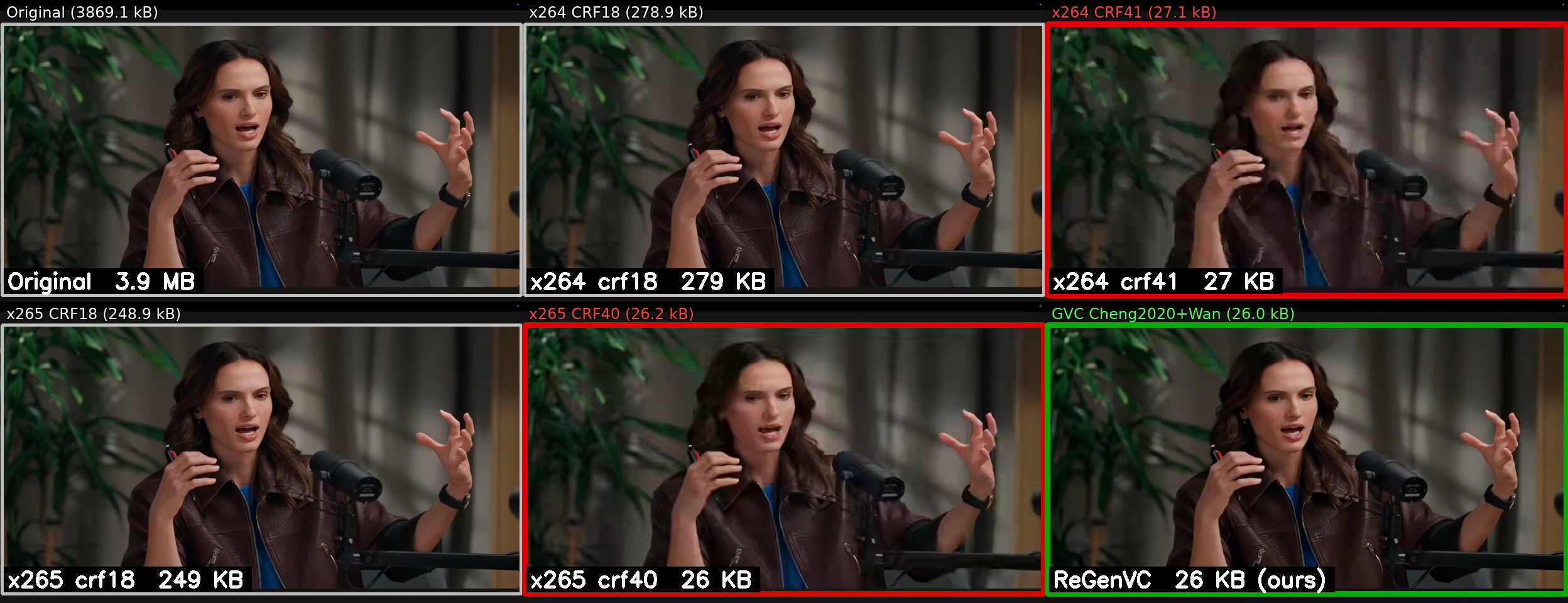}
  \caption{Qualitative comparison on a talking-head frame. \emph{Top:} original,
  x264 crf~18, x264 crf~41. \emph{Bottom:} x265 crf~18, x265 crf~40, ReGenVC
  (ours). At high quality x264/x265 need $\sim$250--280\,kB (crf~18); at
  ultra-low bitrate ($\sim$\SI{26}{kB}, crf~40/41, \textcolor{red}{red boxes})
  they collapse into blocking, whereas ReGenVC
  (\textcolor{green!55!black}{green box}) stays sharp at the same
  $\sim$\SI{26}{kB}.}
  \label{fig:qualitative}
\end{figure*}

Table~\ref{tab:compress} compares ReGenVC against x264 and x265 (FFmpeg) on the
77-frame talking-head test sequence. Two regimes matter. At \emph{high} quality
(crf~18) the traditional codecs need $\sim$250--280\,kB to reach PSNR
$\sim$49\,dB (essentially artifact-free); ReGenVC transmits the same content
in $\sim$\SI{26}{kB}---an \textbf{order of magnitude smaller}. We do not claim
pixel-fidelity parity with such high-rate x264/x265---the reconstruction is
generative---and only assert that at matched ultra-low bitrate ReGenVC remains
sharp while the traditional codecs collapse. At a
\emph{matched} ultra-low bitrate ($\sim$26\,kB, crf~40--41) the traditional codecs
enter their failure mode: PSNR collapses to $\sim$36\,dB and the frames show
severe $8\times8$ blocking. On the comparison frame (Fig.~\ref{fig:qualitative}), the two low-rate traditional panels at matched bitrate---x264 crf~41 and x265 crf~40, both $\sim$\SI{26}{kB}---show pronounced blocking and loss of high-frequency detail, whereas ReGenVC at that bitrate regenerates sharp facial detail from its prior. We report the visual comparison rather than pixel-fidelity summary statistics, which are not meaningful for a generative reconstruction.

\begin{table}[t]
  \centering\small
  \setlength{\tabcolsep}{3pt}
  \caption{Compression on the 77-frame talking-head clip. ReGenVC transmits the clip
  in $\sim$$10\times$ less bitstream than x264/x265 need for essentially
  artifact-free reconstruction (PSNR $\sim$49\,dB); at matched $\sim$\SI{26}{kB}
  the traditional codecs collapse into blocking while ReGenVC stays sharp. PSNR is a
  pixel-fidelity metric and is not the right yardstick for a generative
  reconstruction (see text).}
  \label{tab:compress}
  \begin{tabular}{@{}lccl@{}}
    \toprule
    method & bitstream (kB) & PSNR (dB) & regime \\
    \midrule
    x264 (crf~18) & 278.9 & 48.6 & high-rate \\
    x265 (crf~18) & 249.0 & 49.0 & high-rate \\
    \midrule
    x264 (crf~41) & 27.1 & 36.5 & blocking \\
    x265 (crf~40) & 26.3 & 37.0 & blocking \\
    \textbf{ReGenVC (ours)} & \textbf{26.0}\rlap{$^\dagger$} & --- & \textbf{sharp (gen.)} \\
    \bottomrule
  \end{tabular}\\[2pt]
  {\footnotesize $^\dagger$ $=10.0$ (reference, \texttt{cheng2020\_attn}) $+15.8$ (pose) $+0.2$ (meta).}
\end{table}

\begin{figure*}[t]
  \centering
  \includegraphics[width=\linewidth]{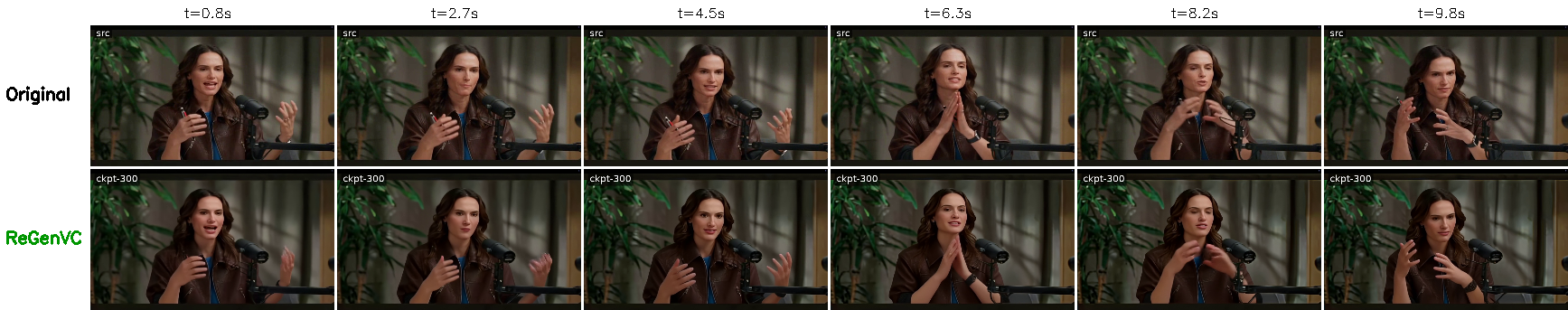}
  \caption{Reconstruction fidelity across the $\sim$10\,s (240-frame) demo
  clip (Sec.~\ref{sec:setup}). \emph{Top row:} original input frames;
  \emph{bottom row:} ReGenVC reconstructions decoded from the
  $\sim$\SI{60}{kB} bitstream, at six time instants spanning the sequence
  (timestamps on top). ReGenVC visibly preserves recognizable identity and
  facial detail across the tested clip.}
  \label{fig:recon}
\end{figure*}

This qualitative advantage is sustained across time on the longer 240-frame
($\sim$10\,s, $\sim$\SI{60}{kB}) clip (Fig.~\ref{fig:recon}): ReGenVC preserves
recognizable identity and sharp facial detail throughout the sequence, while
x264/x265 at the matched ultra-low bitrate remain in the blocking regime shown
above.

\subsection{End-to-end real-time performance}
\begin{figure}[t]
  \centering
  \includegraphics[width=\linewidth]{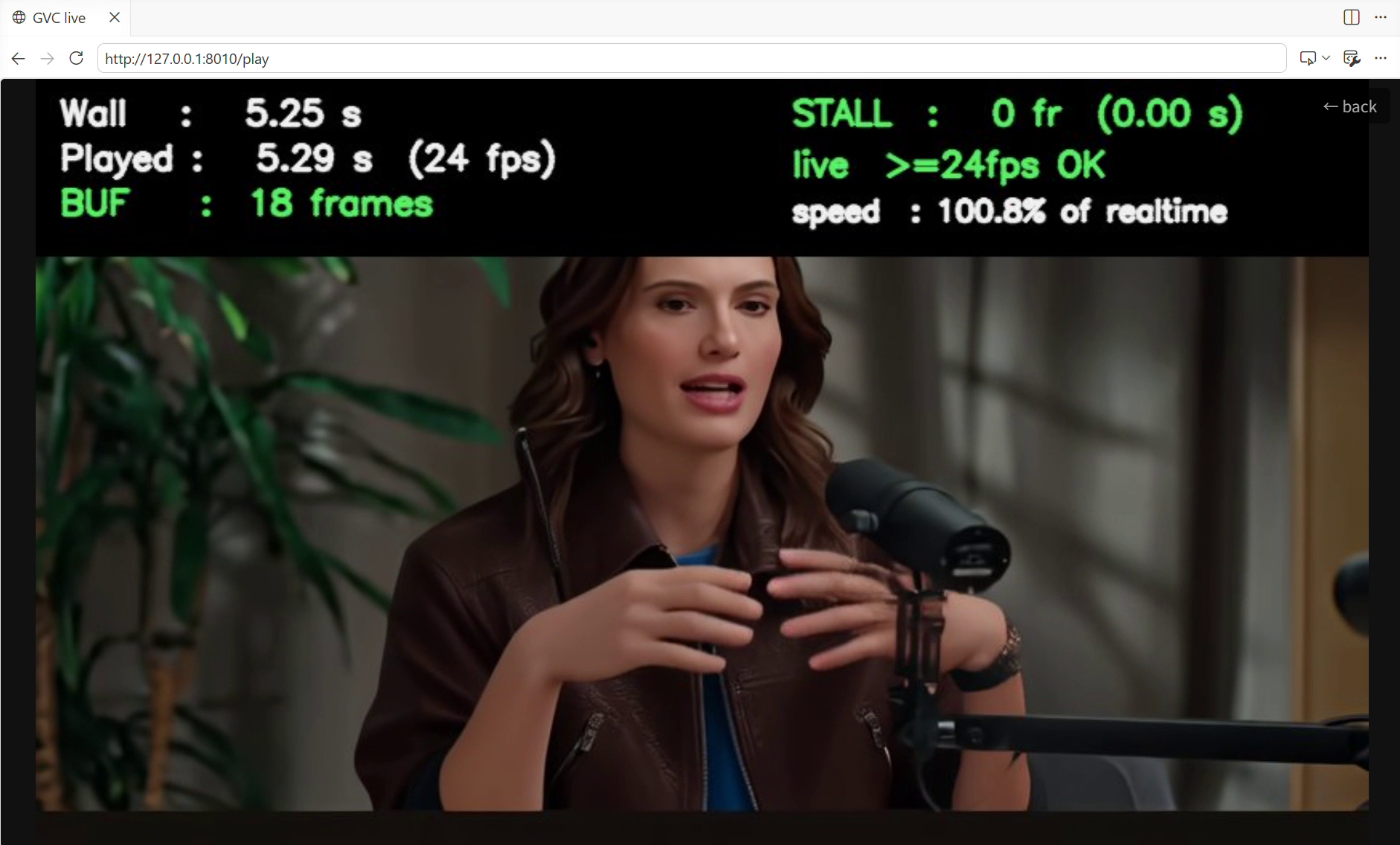}
  \caption{Live browser demo: ReGenVC decoding and streaming at \SI{24}{fps} on
  $8\times$RTX~5090D.}
  \label{fig:demo}
\end{figure}

Table~\ref{tab:e2e} reports the steady-state per-window wall-clock time and its
decomposition under Eq.~\eqref{eq:wall}. The full system sustains
$\wall \approx \SI{972}{ms}$ per window, inside the
\SI{1000}{ms} budget with $\approx\SI{28}{ms}$ ($2.8\%$) of headroom. In the
tested run the system drove a browser video stream at \SI{24}{fps} with no
observed frame underruns (Fig.~\ref{fig:demo}); the narrow headroom leaves the
served pipeline sensitive to per-window jitter, and long-horizon stress testing
is left to future work.

\begin{table}[t]
  \centering\small
  \setlength{\tabcolsep}{5pt}
  \caption{Steady-state per-window decomposition (ms) on $8\times$RTX~5090D,
  full system, reported as the mean over the steady-state phase of the tested
  run. $\wall$ fits inside $\real=\SI{1000}{ms}$ with $\sim$\SI{28}{ms}
  headroom; the anchor-encode $a$ runs serially after $v$ on \texttt{dec\_stream}, so the decode branch pays $v{+}a$; the prep term $p$ and host post-processing $\text{cpu}$ hide under the next window. Terms follow Sec.~\ref{sec:timing}.}
  \label{tab:e2e}
  \begin{tabular}{@{}lcccccc@{}}
    \toprule
    \multicolumn{2}{c}{diffusion path} & \multicolumn{4}{c}{overlapped VAE path} & total \\
    \cmidrule(lr){1-2}\cmidrule(lr){3-6}\cmidrule(l){7-7}
    $w_d$ & $d$ & $v$ & $a$ & $e$ & $w_e$ & $\wall$ \\
    \midrule
    $\sim$78 & $\sim$435 & $\sim$399 & $\sim$60 & $\sim$301 & $\sim$157 & $\approx$\textbf{972} \\
    \bottomrule
  \end{tabular}
\end{table}

The decomposition confirms the model: the diffusion path contributes
$w_d + d \approx \SI{513}{ms}$ and the overlapped VAE path
$\max(v{+}a,\, w_e{+}e) \approx \SI{459}{ms}$ (with decode branch
$v{+}a \approx \SI{459}{ms}$ and concurrent encode branch
$w_e{+}e \approx \SI{458}{ms}$, balanced by design); the two paths sum to
$\approx\SI{972}{ms}$, matching the steady-state $\wall$ and confirming the
Eq.~\eqref{eq:wall} decomposition.

\subsection{Where the latency comes from and which mechanisms are load-bearing}
\label{sec:micro}
The real-time barrier is intrinsic to diffusion decoding, not an artifact of our
design: even a four-step distilled sampler is heavy at video resolution. We
profile each kernel in isolation on a single RTX~5090D in \texttt{bfloat16}
(\SI{10}{iters}, $2$ warm-up); Table~\ref{tab:micro} reports the results. A
single transformer forward costs \SI{368.8}{ms}, so the four-step sampler alone
is $\approx\SI{1475}{ms}$---already $1.5\times$ the frame budget---while the VAE
decode (\SI{957}{ms}) and encode (\SI{564}{ms}) each rival several transformer
forwards. These are fixed costs that \emph{any} diffusion-based codec must pay;
our contribution is to \emph{absorb} them within the budget---USP distributes
each forward and the spatial split parallelizes the VAE, turning a
$\sim$\SI{1475}{ms} serial workload into a real-time pipeline.

\begin{table}[t]
  \centering\small
  \setlength{\tabcolsep}{5pt}
  \caption{Single-GPU \texttt{bf16} kernel latency (mean, ms) on one RTX~5090D.
  The four-step transformer alone exceeds the real-time budget, while each VAE
  pass consumes most of it, motivating multi-GPU sharding and spatial splitting.}
  \label{tab:micro}
  \begin{tabular}{@{}lcc@{}}
    \toprule
    kernel (1 GPU, bf16) & mean (ms) & vs.\ budget \\
    \midrule
    Transformer forward ($1\times$) & 368.8 & $0.37\real$ \\
    Transformer forward ($4\times$, 4-step) & 1475.2 & $1.48\real$ \\
    VAE decode & 957.4 & $0.96\real$ \\
    VAE encode & 564.2 & $0.56\real$ \\
    \bottomrule
  \end{tabular}
\end{table}

\paragraph{Effect of splitting.}
Under the $8$-GPU spatial split the decode drops to $v\approx\SI{399}{ms}$
(Table~\ref{tab:e2e}), a $957.4/399 \approx 2.4\times$ speed-up over the single
device---sub-linear in device count due to halo overlap and gather
communication, but sufficient to bring the VAE path inside the overlapped budget.

\paragraph{Overlap is structurally required.}
By Eq.~\eqref{eq:wall}, disabling cross-window overlap collapses
$\max(v{+}a,\, w_e{+}e)$ into the sum $(v{+}a)+(w_e{+}e)$. With the split VAE
numbers above ($v\approx\SI{399}{ms}$, $e\approx\SI{458}{ms}$, plus the anchor
encode $a$), that puts $\sim\SI{857}{ms}$ of VAE-side cost on top of the
$w_d{+}d$ path---structurally busting the \SI{1000}{ms} budget without ever
running the un-overlapped configuration.

\paragraph{Compilation ablation.}
One further measured optimization matters at the margin: switching from eager
execution to \texttt{torch.compile(mode=default)} reduces \wall{} from
\SI{1005}{ms} to \SI{972}{ms}, a \SI{33}{ms} ($3.3\%$) improvement entirely in
the launch-overhead term $w_d$ (Sec.~\ref{sec:compile})---just enough to carry
the pipeline across the real-time line. The aggressive \texttt{reduce-overhead}
(CUDA-graph) mode is not usable under our cross-stream schedule and is excluded.

\subsection{Discussion and limitations}
\textbf{Domain.} ReGenVC targets pose-drivable talking-head/human video, where a
compact pose signal can drive generation; it is \emph{not} a general-purpose
codec and would not apply to arbitrary scenes without a suitable conditioning
modality. Within this domain, the current setup assumes a fixed camera and a single subject; behavior under multi-person scenes and on streams longer than the tested clip is left for future evaluation. \textbf{Evaluation scope.} Our quantitative compression measurements (Sec.~\ref{sec:compress}) are on a single 77-frame talking-head clip, and sustained multi-window decoding and reconstruction fidelity (Figs.~\ref{fig:demo},~\ref{fig:recon}) are demonstrated on a single 240-frame ($\sim$10\,s) clip of the same type; a benchmark-scale evaluation across a talking-head dataset is left for future work. \textbf{Metric.} Because the reconstruction is generative, PSNR
understates quality; the honest comparison is perceptual, and a full
LPIPS/FVD/identity-consistency study against neural and generative codecs is
left to future work. Perceptual equivalence between the 4-step student and the 20-step DDIM teacher likewise depends on the quality of the distillation and is not claimed here. \textbf{Cost.} Real time here requires an $8$-GPU node: relative to
single-GPU quantized/sparsified decoders we trade higher hardware cost for
throughput \emph{and} model-preserving decoding, and the architectural design is amenable to scaling to larger models or
resolutions by adding devices (though such scaling is not validated in this work).

%% file: sec/6_conclusion.tex
\section{Conclusion}
\label{sec:conclusion}

We presented \textbf{ReGenVC}, an end-to-end generative video codec whose
distilled diffusion decoder meets a strict 1-second per-window real-time
budget on 8 consumer-class GPUs. Its central design choice is
\emph{model-preserving}: rather than shrinking the transformer via
quantization or architectural rewrites, we distribute a single 4-step student
across eight devices via unified sequence parallelism, spatial VAE splitting,
and overlapped cross-window scheduling. Our contributions span an end-to-end
coding scheme, a DMD2 + $x_0$-regression distillation recipe, a
model-preserving multi-GPU decoder schedule, an analytical timing model, and
a hybrid CPU--GPU deployment.

We see four avenues for follow-up work: \emph{(i)} a full
rate--distortion--perception evaluation (LPIPS, FVD, identity consistency)
across a talking-head benchmark; \emph{(ii)} extending beyond talking-head to
general pose- or motion-drivable video; \emph{(iii)} exploring the
resource--quality Pareto frontier---higher reconstruction quality within the
same 8-GPU real-time budget, or the same \wall{} target on fewer GPUs; and
\emph{(iv)} scaling out to multi-node clusters to support larger backbones or
higher resolutions. We hope these directions help move generative video
coding from a largely offline research topic toward practical deployment in
bandwidth-constrained applications such as live streaming, video messaging,
and telepresence.

%% file: main.bbl
\begin{thebibliography}{43}
\providecommand{\natexlab}[1]{#1}
\providecommand{\url}[1]{\texttt{#1}}
\expandafter\ifx\csname urlstyle\endcsname\relax
  \providecommand{\doi}[1]{doi: #1}\else
  \providecommand{\doi}{doi: \begingroup \urlstyle{rm}\Url}\fi

\bibitem[B{\'e}gaint et~al.(2020)B{\'e}gaint, Racap{\'e}, Feltman, and
  Pushparaja]{begaint2020compressai}
Jean B{\'e}gaint, Fabien Racap{\'e}, Simon Feltman, and Akshay Pushparaja.
\newblock Compressai: a pytorch library and evaluation platform for end-to-end
  compression research.
\newblock \emph{arXiv preprint arXiv:2011.03029}, 2020.

\bibitem[Blattmann et~al.(2023)Blattmann, Dockhorn, Kulal, Mendelevitch,
  Kilian, Lorenz, Levi, English, Voleti, Letts, Jampani, and
  Rombach]{blattmann2023svd}
Andreas Blattmann, Tim Dockhorn, Sumith Kulal, Daniel Mendelevitch, Maciej
  Kilian, Dominik Lorenz, Yam Levi, Zion English, Vikram Voleti, Adam Letts,
  Varun Jampani, and Robin Rombach.
\newblock Stable video diffusion: Scaling latent video diffusion models to
  large datasets.
\newblock \emph{arXiv preprint arXiv:2311.15127}, 2023.

\bibitem[Chen et~al.(2024)Chen, Chen, Wang, and Ye]{chen2024gfvc_review}
Bolin Chen, Jie Chen, Shiqi Wang, and Yan Ye.
\newblock Generative face video coding techniques and standardization efforts:
  A review.
\newblock In \emph{Data Compression Conference (DCC)}, 2024.

\bibitem[Cheng et~al.(2020)Cheng, Sun, Takeuchi, and Katto]{cheng2020}
Zhengxue Cheng, Heming Sun, Masaru Takeuchi, and Jiro Katto.
\newblock Learned image compression with discretized gaussian mixture
  likelihoods and attention modules.
\newblock In \emph{IEEE/CVF Conference on Computer Vision and Pattern
  Recognition (CVPR)}, 2020.

\bibitem[Conneau et~al.(2020)Conneau, Khandelwal, Goyal, Chaudhary, Wenzek,
  Guzm{\'a}n, Grave, Ott, Zettlemoyer, and Stoyanov]{conneau2020xlmr}
Alexis Conneau, Kartikay Khandelwal, Naman Goyal, Vishrav Chaudhary, Guillaume
  Wenzek, Francisco Guzm{\'a}n, Edouard Grave, Myle Ott, Luke Zettlemoyer, and
  Veselin Stoyanov.
\newblock Unsupervised cross-lingual representation learning at scale.
\newblock In \emph{Annual Meeting of the Association for Computational
  Linguistics (ACL)}, 2020.

\bibitem[Dao(2024)]{dao2023flashattention2}
Tri Dao.
\newblock Flashattention-2: Faster attention with better parallelism and work
  partitioning.
\newblock In \emph{International Conference on Learning Representations
  (ICLR)}, 2024.

\bibitem[Dao et~al.(2022)Dao, Fu, Ermon, Rudra, and
  R\'e]{dao2022flashattention}
Tri Dao, Daniel~Y. Fu, Stefano Ermon, Atri Rudra, and Christopher R\'e.
\newblock Flashattention: Fast and memory-efficient exact attention with
  io-awareness.
\newblock In \emph{Advances in Neural Information Processing Systems
  (NeurIPS)}, 2022.

\bibitem[Fang et~al.(2024)Fang, Pan, Sun, Li, and Wang]{fang2024xdit}
Jiarui Fang, Jinzhe Pan, Xibo Sun, Aoyu Li, and Jiannan Wang.
\newblock xdit: An inference engine for diffusion transformers (dits) with
  massive parallelism.
\newblock \emph{arXiv preprint arXiv:2411.01738}, 2024.

\bibitem[Fang et~al.(2025)Fang, Pan, Li, Sun, and Wang]{wang2024pipefusion}
Jiarui Fang, Jinzhe Pan, Aoyu Li, Xibo Sun, and Jiannan Wang.
\newblock Pipefusion: Patch-level pipeline parallelism for diffusion
  transformers inference.
\newblock In \emph{Advances in Neural Information Processing Systems
  (NeurIPS)}, 2025.

\bibitem[Gao et~al.(2026)Gao, Pan, Liu, Li, Chen, and Tian]{gao2026zerogvc}
Yixin Gao, Xiaohan Pan, Lin Liu, Xin Li, Zhibo Chen, and Qi Tian.
\newblock {ZeroGVC}: Zero-shot generative video compression with autoregressive
  diffusion priors.
\newblock \emph{arXiv preprint arXiv:2606.22371}, 2026.

\bibitem[Ho et~al.(2020)Ho, Jain, and Abbeel]{ho2020ddpm}
Jonathan Ho, Ajay Jain, and Pieter Abbeel.
\newblock Denoising diffusion probabilistic models.
\newblock In \emph{Advances in Neural Information Processing Systems
  (NeurIPS)}, 2020.

\bibitem[Huang et~al.(2025)Huang, Li, He, Zhou, and
  Shechtman]{huang2025selfforcing}
Xun Huang, Zhengqi Li, Guande He, Mingyuan Zhou, and Eli Shechtman.
\newblock Self forcing: Bridging the train-test gap in autoregressive video
  diffusion.
\newblock In \emph{Advances in Neural Information Processing Systems
  (NeurIPS)}, 2025.

\bibitem[Ilharco et~al.(2021)Ilharco, Wortsman, Wightman, Gordon, Carlini,
  Taori, Dave, Shankar, Namkoong, Miller, Hajishirzi, Farhadi, and
  Schmidt]{ilharco2021openclip}
Gabriel Ilharco, Mitchell Wortsman, Ross Wightman, Cade Gordon, Nicholas
  Carlini, Rohan Taori, Achal Dave, Vaishaal Shankar, Hongseok Namkoong, John
  Miller, Hannaneh Hajishirzi, Ali Farhadi, and Ludwig Schmidt.
\newblock Openclip.
\newblock \url{https://doi.org/10.5281/zenodo.5143773}, 2021.

\bibitem[Jacobs et~al.(2023)Jacobs, Tanaka, Zhang, Zhang, Song, Rajbhandari,
  and He]{jacobs2023ulysses}
Sam~Ade Jacobs, Masahiro Tanaka, Chengming Zhang, Minjia Zhang, Shuaiwen~Leon
  Song, Samyam Rajbhandari, and Yuxiong He.
\newblock Deepspeed ulysses: System optimizations for enabling training of
  extreme long sequence transformer models.
\newblock \emph{arXiv preprint arXiv:2309.14509}, 2023.

\bibitem[Kodaira et~al.(2025{\natexlab{a}})Kodaira, Hou, Hou, Georgopoulos,
  Juefei-Xu, Tomizuka, and Zhao]{kodaira2025streamdit}
Akio Kodaira, Tingbo Hou, Ji Hou, Markos Georgopoulos, Felix Juefei-Xu,
  Masayoshi Tomizuka, and Yue Zhao.
\newblock Streamdit: Real-time streaming text-to-video generation.
\newblock \emph{arXiv preprint arXiv:2507.03745}, 2025{\natexlab{a}}.

\bibitem[Kodaira et~al.(2025{\natexlab{b}})Kodaira, Xu, Hazama, Yoshimoto,
  Ohno, Mitsuhori, Sugano, Cho, Liu, Tomizuka, and
  Keutzer]{kodaira2024streamdiffusion}
Akio Kodaira, Chenfeng Xu, Toshiki Hazama, Takanori Yoshimoto, Kohei Ohno,
  Shogo Mitsuhori, Soichi Sugano, Hanying Cho, Zhijian Liu, Masayoshi Tomizuka,
  and Kurt Keutzer.
\newblock Streamdiffusion: A pipeline-level solution for real-time interactive
  generation.
\newblock In \emph{IEEE/CVF International Conference on Computer Vision
  (ICCV)}, 2025{\natexlab{b}}.

\bibitem[Li et~al.(2024)Li, Li, and Lu]{li2024dcvcfm}
Jiahao Li, Bin Li, and Yan Lu.
\newblock Neural video compression with feature modulation.
\newblock In \emph{IEEE/CVF Conference on Computer Vision and Pattern
  Recognition (CVPR)}, 2024.

\bibitem[Ling et~al.(2026)Ling, Zhou, Li, Chen, Tian, Lu, and
  Zhang]{ling2026freegvc}
Xiaoyue Ling, Chuqin Zhou, Chunyi Li, Yunuo Chen, Yuan Tian, Guo Lu, and Wenjun
  Zhang.
\newblock {Free-GVC}: Towards training-free extreme generative video
  compression with temporal coherence.
\newblock \emph{arXiv preprint arXiv:2602.09868}, 2026.

\bibitem[Lipman et~al.(2023)Lipman, Chen, Ben-Hamu, Nickel, and
  Le]{lipman2023flowmatching}
Yaron Lipman, Ricky T.~Q. Chen, Heli Ben-Hamu, Maximilian Nickel, and Matt Le.
\newblock Flow matching for generative modeling.
\newblock In \emph{International Conference on Learning Representations
  (ICLR)}, 2023.

\bibitem[Liu et~al.(2023)Liu, Zaharia, and Abbeel]{liu2023ringattention}
Hao Liu, Matei Zaharia, and Pieter Abbeel.
\newblock Ring attention with blockwise transformers for near-infinite context.
\newblock \emph{arXiv preprint arXiv:2310.01889}, 2023.

\bibitem[Liu et~al.(2024)Liu, Zhang, Ma, Peng, and Liu]{liu2023instaflow}
Xingchao Liu, Xiwen Zhang, Jianzhu Ma, Jian Peng, and Qiang Liu.
\newblock Instaflow: One step is enough for high-quality diffusion-based
  text-to-image generation.
\newblock In \emph{International Conference on Learning Representations
  (ICLR)}, 2024.

\bibitem[Luo et~al.(2023)Luo, Tan, Huang, Li, and Zhao]{luo2023lcm}
Simian Luo, Yiqin Tan, Longbo Huang, Jian Li, and Hang Zhao.
\newblock Latent consistency models: Synthesizing high-resolution images with
  few-step inference.
\newblock \emph{arXiv preprint arXiv:2310.04378}, 2023.

\bibitem[Radford et~al.(2021)Radford, Kim, Hallacy, Ramesh, Goh, Agarwal,
  Sastry, Askell, Mishkin, Clark, Krueger, and Sutskever]{radford2021clip}
Alec Radford, Jong~Wook Kim, Chris Hallacy, Aditya Ramesh, Gabriel Goh,
  Sandhini Agarwal, Girish Sastry, Amanda Askell, Pamela Mishkin, Jack Clark,
  Gretchen Krueger, and Ilya Sutskever.
\newblock Learning transferable visual models from natural language
  supervision.
\newblock In \emph{International Conference on Machine Learning (ICML)}, 2021.

\bibitem[Raffel et~al.(2020)Raffel, Shazeer, Roberts, Lee, Narang, Matena,
  Zhou, Li, and Liu]{raffel2020t5}
Colin Raffel, Noam Shazeer, Adam Roberts, Katherine Lee, Sharan Narang, Michael
  Matena, Yanqi Zhou, Wei Li, and Peter~J. Liu.
\newblock Exploring the limits of transfer learning with a unified text-to-text
  transformer.
\newblock \emph{Journal of Machine Learning Research (JMLR)}, 2020.

\bibitem[Rombach et~al.(2022)Rombach, Blattmann, Lorenz, Esser, and
  Ommer]{rombach2022ldm}
Robin Rombach, Andreas Blattmann, Dominik Lorenz, Patrick Esser, and Bj\"orn
  Ommer.
\newblock High-resolution image synthesis with latent diffusion models.
\newblock In \emph{IEEE/CVF Conference on Computer Vision and Pattern
  Recognition (CVPR)}, 2022.

\bibitem[Salimans and Ho(2022)]{salimans2022progressive}
Tim Salimans and Jonathan Ho.
\newblock Progressive distillation for fast sampling of diffusion models.
\newblock In \emph{International Conference on Learning Representations
  (ICLR)}, 2022.

\bibitem[Siarohin et~al.(2019)Siarohin, Lathuili{\`e}re, Tulyakov, Ricci, and
  Sebe]{siarohin2019fomm}
Aliaksandr Siarohin, St{\'e}phane Lathuili{\`e}re, Sergey Tulyakov, Elisa
  Ricci, and Nicu Sebe.
\newblock First order motion model for image animation.
\newblock In \emph{Advances in Neural Information Processing Systems
  (NeurIPS)}, 2019.

\bibitem[Song et~al.(2023)Song, Dhariwal, Chen, and
  Sutskever]{song2023consistency}
Yang Song, Prafulla Dhariwal, Mark Chen, and Ilya Sutskever.
\newblock Consistency models.
\newblock In \emph{International Conference on Machine Learning (ICML)}, 2023.

\bibitem[Wan et~al.(2025)Wan, Zheng, and Fan]{wan2024m3cvc}
Rui Wan, Qi Zheng, and Yibo Fan.
\newblock {M3-CVC}: Controllable video compression with multimodal generative
  models.
\newblock In \emph{IEEE International Conference on Acoustics, Speech and
  Signal Processing (ICASSP)}, 2025.

\bibitem[{Wan Team} et~al.(2025)]{wan2025}
{Wan Team} et~al.
\newblock Wan: Open and advanced large-scale video generative models.
\newblock \emph{arXiv preprint arXiv:2503.20314}, 2025.

\bibitem[Wang et~al.(2025)Wang, Su, Kothandaraman, Huang, Hajiesmaili, and
  Sitaraman]{wang2025disco}
Lingdong Wang, Guan-Ming Su, Divya Kothandaraman, Tsung-Wei Huang, Mohammad
  Hajiesmaili, and Ramesh~K. Sitaraman.
\newblock Low-bitrate video compression through semantic-conditioned diffusion.
\newblock \emph{arXiv preprint arXiv:2512.00408}, 2025.

\bibitem[Wang et~al.(2021)Wang, Mallya, and Liu]{wang2021face_vid2vid}
Ting-Chun Wang, Arun Mallya, and Ming-Yu Liu.
\newblock One-shot free-view neural talking-head synthesis for video
  conferencing.
\newblock In \emph{IEEE/CVF Conference on Computer Vision and Pattern
  Recognition (CVPR)}, 2021.

\bibitem[Wang et~al.(2026)Wang, Man, Li, Wang, Fan, and Zhao]{wang2025tgvc}
Zhitao Wang, Hengyu Man, Wenrui Li, Xingtao Wang, Xiaopeng Fan, and Debin Zhao.
\newblock {T-GVC}: Trajectory-guided generative video coding at ultra-low
  bitrates.
\newblock In \emph{AAAI Conference on Artificial Intelligence (AAAI)}, 2026.

\bibitem[Wu et~al.(2026)Wu, Chen, Lu, Song, Yang, Tao, and Zhang]{wu2026jscgc}
Tong Wu, Zhiyong Chen, Guo Lu, Li Song, Feng Yang, Meixia Tao, and Wenjun
  Zhang.
\newblock Joint source-channel-generation coding: From distortion-oriented
  reconstruction to semantic-consistent generation.
\newblock \emph{arXiv preprint arXiv:2601.12808}, 2026.

\bibitem[Yang et~al.(2023)Yang, Zeng, Yuan, and Li]{yang2023dwpose}
Zhendong Yang, Ailing Zeng, Chun Yuan, and Yu Li.
\newblock Effective whole-body pose estimation with two-stages distillation.
\newblock In \emph{IEEE/CVF International Conference on Computer Vision
  Workshops (ICCVW)}, 2023.

\bibitem[Yang et~al.(2025)Yang, Teng, Zheng, Ding, Huang, Xu, Yang, Hong,
  Zhang, Feng, Yin, Zhang, Wang, Cheng, Xu, Gu, Dong, and
  Tang]{yang2024cogvideox}
Zhuoyi Yang, Jiayan Teng, Wendi Zheng, Ming Ding, Shiyu Huang, Jiazheng Xu,
  Yuanming Yang, Wenyi Hong, Xiaohan Zhang, Guanyu Feng, Da Yin, Yuxuan Zhang,
  Weihan Wang, Yean Cheng, Bin Xu, Xiaotao Gu, Yuxiao Dong, and Jie Tang.
\newblock Cogvideox: Text-to-video diffusion models with an expert transformer.
\newblock In \emph{International Conference on Learning Representations
  (ICLR)}, 2025.

\bibitem[Yi et~al.(2025)Yi, Xu, Shao, Zhang, and Li]{yi2025conditional}
Fangqiu Yi, Jingyu Xu, Jiawei Shao, Chi Zhang, and Xuelong Li.
\newblock Conditional video generation for high-efficiency video compression.
\newblock \emph{arXiv preprint arXiv:2507.15269}, 2025.

\bibitem[Yin et~al.(2024{\natexlab{a}})Yin, Gharbi, Park, Zhang, Shechtman,
  Durand, and Freeman]{yin2024dmd2}
Tianwei Yin, Micha\"el Gharbi, Taesung Park, Richard Zhang, Eli Shechtman,
  Fr\'edo Durand, and William~T. Freeman.
\newblock Improved distribution matching distillation for fast image synthesis.
\newblock \emph{Advances in Neural Information Processing Systems (NeurIPS)},
  2024{\natexlab{a}}.

\bibitem[Yin et~al.(2024{\natexlab{b}})Yin, Gharbi, Zhang, Shechtman, Durand,
  Freeman, and Park]{yin2024dmd}
Tianwei Yin, Micha\"el Gharbi, Richard Zhang, Eli Shechtman, Fr\'edo Durand,
  William~T. Freeman, and Taesung Park.
\newblock One-step diffusion with distribution matching distillation.
\newblock In \emph{IEEE/CVF Conference on Computer Vision and Pattern
  Recognition (CVPR)}, 2024{\natexlab{b}}.

\bibitem[Yin et~al.(2025)Yin, Zhang, Zhang, Freeman, Durand, Shechtman, and
  Huang]{yin2025causvid}
Tianwei Yin, Qiang Zhang, Richard Zhang, William~T. Freeman, Fr\'edo Durand,
  Eli Shechtman, and Xun Huang.
\newblock From slow bidirectional to fast autoregressive video diffusion
  models.
\newblock \emph{IEEE/CVF Conference on Computer Vision and Pattern Recognition
  (CVPR)}, 2025.

\bibitem[Zhang et~al.(2023)Zhang, Rao, and Agrawala]{zhang2023controlnet}
Lvmin Zhang, Anyi Rao, and Maneesh Agrawala.
\newblock Adding conditional control to text-to-image diffusion models.
\newblock In \emph{IEEE/CVF International Conference on Computer Vision
  (ICCV)}, 2023.

\bibitem[Zhao et~al.(2026)Zhao, Pan, He, Yu, Chen, Ye, Liu, Xie, and
  Han]{zhao2026sanastreaming}
Yuyang Zhao, Yicheng Pan, Qiyuan He, Jincheng Yu, Junsong Chen, Tian Ye, Haozhe
  Liu, Enze Xie, and Song Han.
\newblock Sana-streaming: Real-time streaming video editing with hybrid
  diffusion transformer.
\newblock \emph{arXiv preprint arXiv:2605.30409}, 2026.

\bibitem[Zhuang et~al.(2025)Zhuang, Guo, Cai, Li, Liu, Yuan, and
  Xue]{zhuang2025flashvsr}
Junhao Zhuang, Shi Guo, Xin Cai, Xiaohui Li, Yihao Liu, Chun Yuan, and Tianfan
  Xue.
\newblock Flashvsr: Towards real-time diffusion-based streaming video
  super-resolution.
\newblock \emph{arXiv preprint arXiv:2510.12747}, 2025.

\end{thebibliography}
